\documentclass[aps,prl,twocolumn,groupedaddress]{revtex4-1}

\usepackage{graphicx} 
\usepackage{epsfig}
\usepackage{amsmath} 
\usepackage{amsthm} 
\usepackage{amssymb}	
\newcommand{\eq}[1]{\begin{align} #1 \end{align}}

\newcommand{\keff}[0]{\ensuremath{k_\text{eff}}}
\usepackage{graphics} 
\usepackage{hyperref} 
\hypersetup{
    colorlinks,
    citecolor=red,
    filecolor=black,
    linkcolor=red,
    urlcolor=blue
}

\begin{document}


\title{Finite-Size Scaling at the Jamming Transition}


\author{Carl P. Goodrich}

\author{Andrea J. Liu}
\email[]{ajliu@physics.upenn.edu}
\affiliation{Department of Physics, University of Pennsylvania, Philadelphia, Pennsylvania 19104, USA}

\author{Sidney R. Nagel}
\affiliation{James Franck Institute, The University of Chicago, Chicago, Illinois 60637, USA}


\date{\today}

\begin{abstract}
We present an analysis of finite-size effects in jammed packings of $N$ soft, frictionless spheres at zero temperature.  There is a $\frac 1N$ correction to the discrete jump in the contact number at the transition so that jammed packings exist \emph{only above} isostaticity. As a result, the canonical power-law scalings of the contact number and elastic moduli break down at low pressure. These quantities exhibit scaling collapse with a non-trivial scaling function, demonstrating that the jamming transition can be considered a phase transition. Scaling is achieved as a function of $N$ in both 2 and 3 dimensions, indicating an upper critical dimension of 2.
%
\end{abstract}

\pacs{}

\maketitle

Numerical simulations of particulate systems are unavoidably limited to a finite number of particles.  It has long been recognized in the context of phase transitions that this limitation can be exploited~\cite{Fisher:1972hr} to yield insight into the nature of the transition.  In the context of the zero-temperature jamming transition of frictionless sphere packings~\cite{Liu:2010jx}, a finite-size analysis should be particularly valuable because the transition appears to be a rare example of a random first-order transition in finite dimensions, characterized by a discontinuous jump in the contact number ({\it i.e.} the average number of interacting neighbors) and power-law scaling~\cite{Durian:1995eo,OHern:2003vq} as well as diverging length scales~\cite{Silbert:2005vw,Wyart:2005wv,Ellenbroek:2006df}.

In this paper, we establish that there are finite-size corrections to the contact number and moduli above the jamming transition. We also reveal novel finite-size behavior close to the transition that can be scaled to collapse onto a single curve, firmly establishing the connection between jamming and phase transitions. While previous work by Olsson and Teitel~\cite{Olsson:2007df} demonstrated scaling collapse in the unjammed regime, our focus is on jammed systems above the transition. We find that all finite-size effects scale with $L^d\sim N$ in $d$ dimensions, where $d=2,3$, $L$ is the linear size of the system, and $N$ is the total number of particles. Such scaling is expected of a system at or above its critical dimension~\cite{BINDER:1985vl} and implies that the jamming transition has an upper critical dimension of 2.  This is consistent with the observation that the power-law exponents are the same in 2 and 3 dimensions~\cite{OHern:2003vq}, as well as an argument that fluctuations should be unimportant for $d \ge 2$~\cite{Wyart:2005jna}.

We consider disordered packings of $N$ frictionless spheres at temperature $T=0$ and pressure, $p$, with a finite-range, repulsive potential between particles $i$ and $j$: 
\eq{V(r_{ij}) = \frac \epsilon \alpha \left(1-r_{ij}/\sigma_{ij}\right)^\alpha	\label{V_of_r}}
only if $r_{ij} \leq \sigma_{ij}$, where $r_{ij}$ is the center-to-center distance, $\sigma_{ij}$ is the sum of their radii, and $\epsilon \equiv 1$ sets the energy scale.  The effective spring constant between contacts is $\keff \equiv \left<\frac{\partial^2 V(r_{ij})}{\partial r_{ij}^2} \right>$~\cite{Liu:2010jx}.  Each packing is relaxed to a local energy minimum.  We then remove the small fraction of ``rattler'' particles that do not contribute to the rigidity of the system~\cite{OHern:2003vq}. 

Before counting constraints for finite systems, we must specify what it means for a system to be jammed. One possible definition is that, in the absence of rattlers, the only zero-frequency vibrational modes are associated with global translation of the system~\cite{OHern:2003vq}. For $N$ spheres in $d$ dimensions, there are $dN$ degrees of freedom and $d$ global translations so that $dN-d$ of the degrees of freedom must be constrained.  This requires that the number of conacts, $N_\text{contact}\ge dN-d$. Since the contact number $Z \equiv \frac {2N_\text{contact}}N$, we obtain
\eq{Z_\text{iso}^N & \equiv 2d - \frac{2d}N. \label{Ziso_N_def}}
This is the contact number required for the system to have no soft modes beyond those corresponding to global translations. In the infinite-size limit, this reduces to the isostatic condition, $Z_\text{iso}^\infty = 2d$, consistent with previous results~\cite{Durian:1995eo,OHern:2003vq}. However, this definition relies on the choice of relevant degrees of freedom. Rattlers, for example, have no effect on the elastic properties of a packing but contribute $d$ zero modes each if not removed. Similarly, a sphere can rotate about its center without any effect on the packing. Thus, this definition can break down, as it does when generalized to packings of ellipsoids~\cite{Donev:2007go,Zeravcic:2009wo,Mailman:2009ct}.  

A more physical requirement is that system have a positive bulk modulus, $B$.  The minimum number of contacts needed for a packing of $N$ spheres to have a positive bulk modulus, $N^\text{min}_\text{contact}$, is
\eq{N_{\text{contact}}^\text{min}= dN -d + 1 \label{Nccount}}
~so that the minimum contact number is:
\eq{Z_B^N & \equiv Z_\text{iso}^N + \frac 2N = 2d - \frac{2d}N + \frac 2N. \label{Z_B_def}
}

In principle, packings with $B>0$ are not forbidden from having complicated soft modes. For sphere packings (with rattlers removed) we have never observed such nontrivial soft modes and therefore assume in the following argument that they do not exist for generic packings.  In this case, at least one {\it additional} contact above $dN-d$ is required for the system to have a positive bulk modulus. 

The origin of this extra contact can be understood by treating the size of the periodic box as a degree of freedom. When $Z\leq Z_\text{iso}^N$, there are at most as many constraint equations as particle degrees of freedom. If there are no nontrivial soft modes, it is possible to satisfy the constraints $r_{ij}=\sigma_{ij}$ for every contact. Thus, by Eq.~\eqref{V_of_r}, the total energy and pressure must be zero. 
Since any deformation in the linear regime does not form any new contacts, the energy remains constant and the bulk modulus, $B$, must be zero. Therefore at least one additional contact is needed for the system to have a positive bulk modulus or pressure. This additional contact corresponds to the last term in Eq.~\eqref{Nccount} and leads to $Z_B^N$ in Eq.~\eqref{Z_B_def}.   For any positive pressure, the contact number should satisfy $Z \ge Z_B^N$, and we  expect that
\eq{\lim_{p \rightarrow 0^+} Z=Z_B^N. \label{Z_approaches_ZBN}	}

A third possible definition of jamming is that the system have a positive shear modulus, $G$, as well as bulk modulus.  Dagois-Bohy, et al.~\cite{DagoisBohy:2012we} have recently shown that packings can be constructed to have positive bulk and shear moduli by allowing the {\it shape} of the box to vary during minimization.  In two dimensions, this introduces 2 extra degrees of freedom for the square box to distort to a rhombus or rectangle.  In $d$ dimensions, there are $\frac 12 d(d+1) -1$ degrees of freedom corresponding to the shape of the box. Therefore, the extension of our counting argument to such ``shear-stabilized" packings predicts a minimum contact number of $Z_{BG}^N \equiv 2d - 2d/N + d(d+1)/N$. This exactly agrees with the findings of Dagois-Bohy, et al.~\cite{DagoisBohy:2012we}.

\begin{figure}
	\centering
	\includegraphics[height=1.4\linewidth,angle=-90,viewport=50 0 490 600, clip]{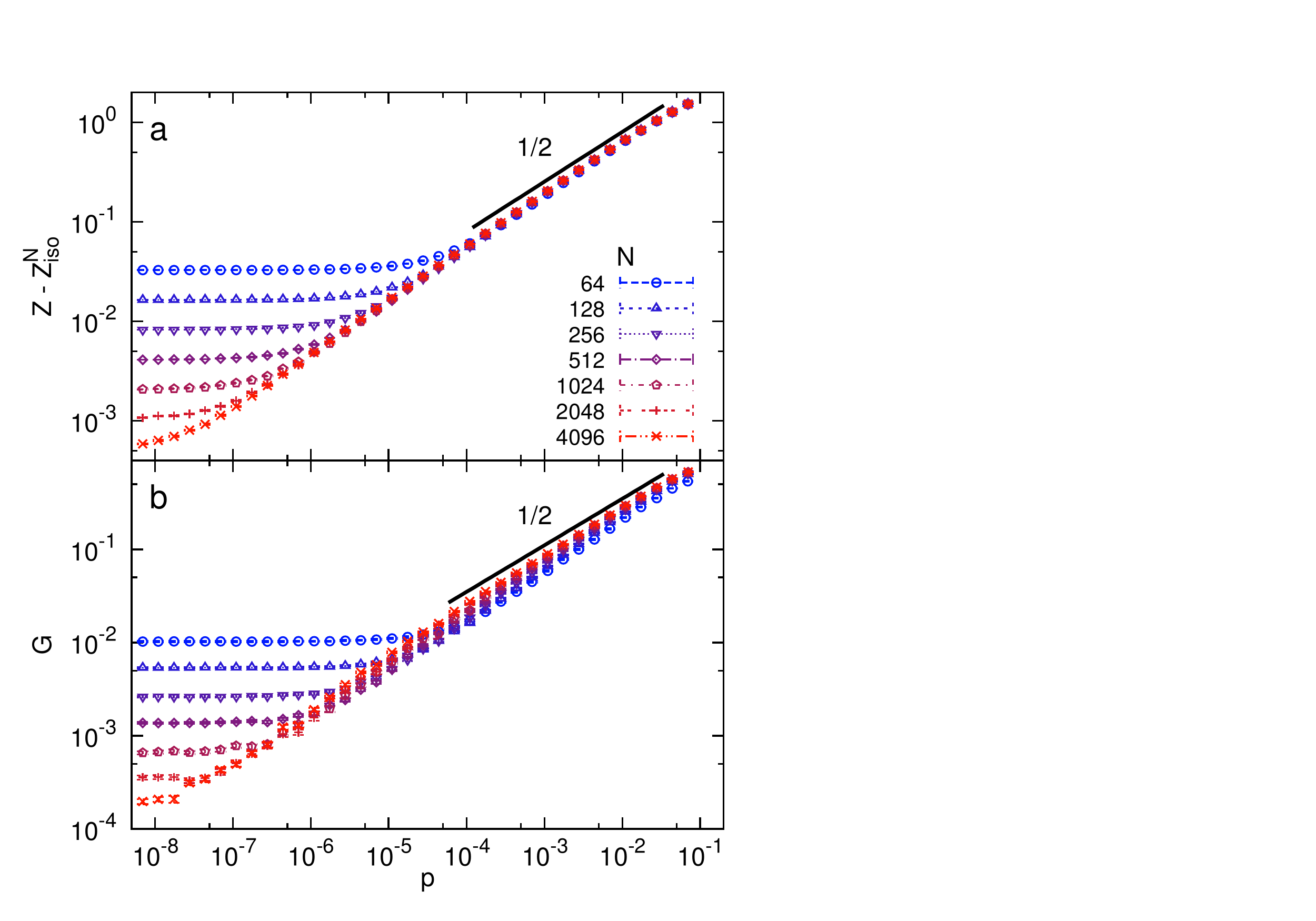}
	\caption{\label{FigureA}(a) $Z - Z_\text{iso}^N$ and (b) $G$ as a function of pressure for different system sizes in 2 dimensions. For both quantities, the power law exponent of $1/2$, observed at high pressures, agrees with the known scaling for harmonic potentials. At low pressures, however, finite-size effects dominate. $G$ is averaged over configurations and shear directions.
	}
\end{figure}

\begin{figure*}
	\centering
	\includegraphics[height=1\linewidth,angle=-90,viewport= 0 12 310 720, clip]{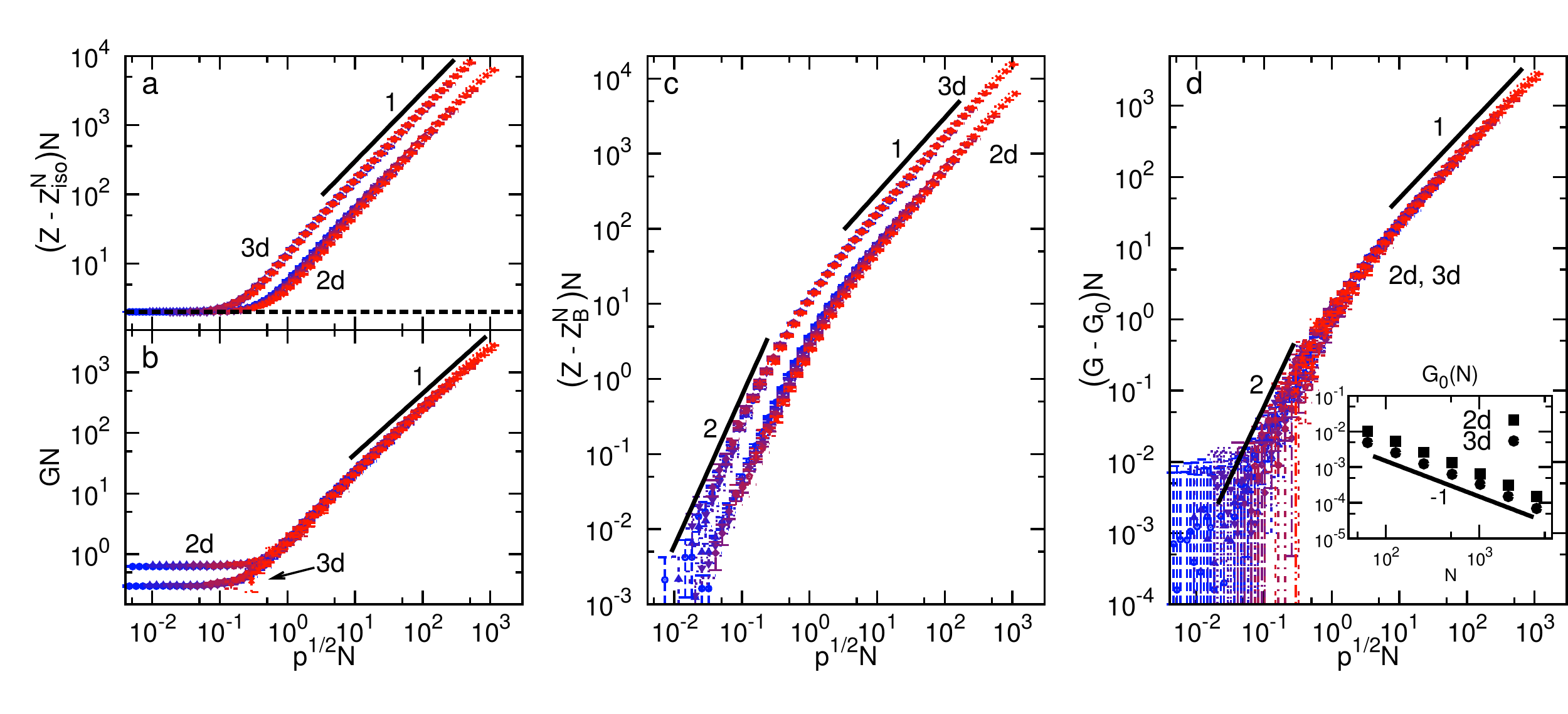}
	\caption{\label{FigureC2} Collapse of (a) $Z - Z_\text{iso}^N$ and (b) $G$ in 2 and 3 dimensions. In the zero pressure limit, $\left(Z - Z_\text{iso}^N\right) N \rightarrow 2$ (dashed line), which corresponds to a single contact above isostaticity. (c) Collapse of $Z - Z_B^N$ in 2 and 3 dimensions. (d) Collapse of $G-G_0$ in 2 and 3 dimensions. The scaling function is qualitatively similar to that of $Z - Z_B^N$. Inset: the plateau value $G_0$ is proportional to $\frac 1N$. Symbols and colors are the same as in Fig.~\ref{FigureA}.}
\end{figure*}

To test the prediction in Eq. (\ref{Z_approaches_ZBN}) and examine finite-size effects, we generated packings of particles with harmonic repulsions given by $V(r_{ij})$ with $\alpha=2$ for systems ranging from $N=64$ to $N=4096$. For this potential, $\keff $ is independent of $r_{ij}$ (and therefore compression) as long as the particles overlap. The relative radii in 2 dimensions were chosen from a flat distribution between $\sigma \equiv 1$ and 1.4 $\sigma$, while in 3 dimensions a bidisperse mixture of ratio 1:1.4 was used.  We fixed the box shape and used pressure as the control variable to produce packings with a positive bulk modulus.  These packings correspond to what Dagois-Bohy, et al.~\cite{DagoisBohy:2012we} refer to as the ``compression-only" ensemble.  

Mechanically stable configurations were generated for a range of pressures spanning 7 orders of magnitude. In a square (cubic) periodic box, particles were placed at random. The system was then quenched to a local energy minimum (using a combination of linesearch methods, Newton's method and the FIRE algorithm~\cite{Bitzek:2006bw} to maximize accuracy and efficiency), and the packing fraction was adjusted until a target pressure was reached. Systems were thrown out if the minimization algorithms did not converge.  For each dimension, system size, and pressure, quantities were averaged over 1000 to 5000 packings. The shear and bulk moduli were calculated via linear response from the dynamical matrix as in \cite{Ellenbroek:2009to,Ellenbroek:2009dp}.  In finite systems, there is a well-defined linear regime in which the contact network is fixed~\cite{Schreck:2011kl}. By using linear response, we ensure that the elastic moduli are calculated in this regime.

\begin{figure}
	\centering
	\includegraphics[height=1.43\linewidth,angle=-90,viewport=200 0 500 600, clip]{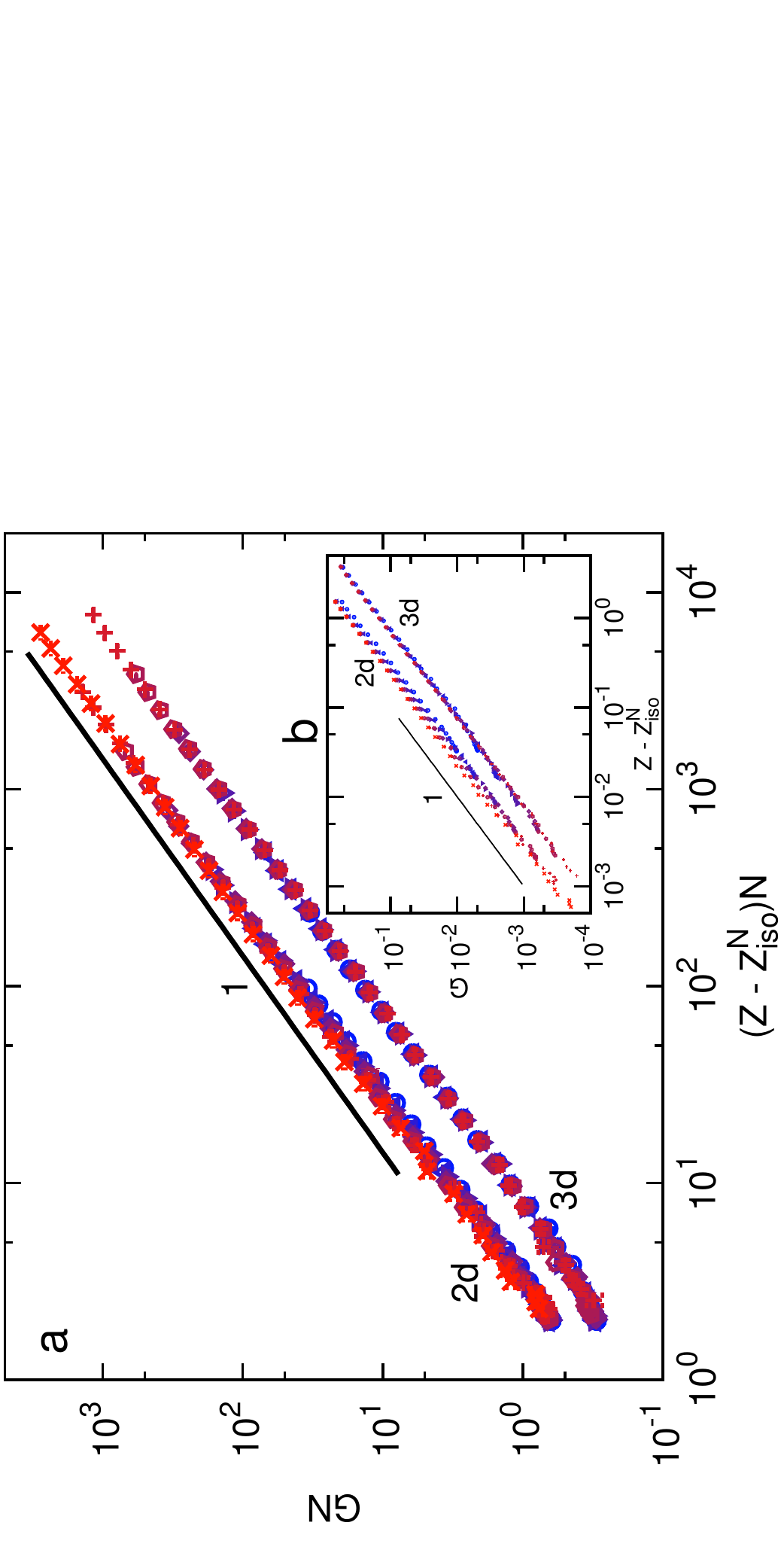}
	\caption{\label{FigureD} $GN \sim \left(Z - Z_\text{iso}^N\right)N$ in both 2 and 3 dimensions. Symbols and colors are the same as in Fig.~\ref{FigureA}.}
\end{figure}

Fig.~\ref{FigureA}(a) and (b) shows both $Z-Z_\text{iso}^N$ and $G$ versus $p$ in 2 dimensions.  
Similar results are obtained in 3 dimensions (see Figs.~\ref{FigureC2} and \ref{FigureD}).
As expected, $Z - Z_\text{iso}^N \sim p^{1/2}$ at high pressures, consistent with previous studies~\cite{Durian:1995eo,OHern:2003vq}, but approaches $2/N$ at low pressures in accord with Eqs.~\eqref{Z_B_def} and~\eqref{Z_approaches_ZBN}.  Thus, one extra contact is needed beyond the isostatic value in order for the bulk modulus to be positive, as predicted.  Moreover, Fig.~\ref{FigureC2}(a) shows that the data collapse when $\left(Z- Z_\text{iso}^N \right) N$, related to the total number of excess contacts, is plotted versus $p^{1/2}N$.  

It is not obvious from constraint-counting arguments alone that at the jamming transition the contact number data should obey finite-size scaling: $Z- Z^N_\text{iso} = N^{y} F(p^{z} N)$ for some $y$ and $z$.   However, if it does then we can show that $y=-1$ and $z=1/2$, consistent with Fig.~\ref{FigureC2}(a).  By counting constraints, we have argued that $Z-Z_\text{iso}^N \rightarrow Z_B^N-Z_\text{iso}^N = 2/N$ as $p \rightarrow 0$.  This is satisfied if $\lim_{x \rightarrow 0}F(x)= 2$ and $y=-1$.  In the large $N$ limit, on the other hand, we must recover the asymptotic scaling relation $Z-Z^N_\text{iso} \sim p^{1/2}$, independent of $N$.  This can only be satisfied if $F(x)\sim x$ at large $x$, and $z=1/2$.  Therefore, the only possible scaling is
\eq{Z- Z^N_\text{iso} = \frac 1N F(p^{1/2} N), \label{Zscalingfunction}}
where $F(x) \sim 1$ for small $x$ and $F(x) \sim x$ for large $x$ (see Fig.~\ref{FigureC2}(a)).  

The shear modulus, shown in Fig.~\ref{FigureA}(b), displays $G\sim p^{1/2}$ at high pressures, again consistent with previous studies \cite{Durian:1995eo,OHern:2003vq}. As the pressure is lowered, however, $G$ develops a plateau that is proportional to $\frac 1N$. For a system of $N$ spheres, one would expect that if one extra contact is required to constrain the size of the periodic box so that $B>0$, additional contacts would be required to constrain the shape of the box as well so that $G>0$, as found by Dagois-Bohy, et al.~\cite{DagoisBohy:2012we}.  However, Fig.~\ref{FigureA}(b) shows that although the shear modulus is not positive in all directions for all configurations, the angle- and configuration-averaged shear modulus is positive with the addition of only one extra contact.  To understand this, note that the shear modulus measures the response to a deformation at constant volume; the {\it size} of the periodic box is held fixed under shear strain and is therefore no longer an independent degree of freedom as it was under compression. This allows the extra contact in Eq.~\eqref{Nccount} to do double duty-- it can contribute to the stability of the system against shear as well as compression.  This extra contact is the origin of the plateau in $G$.

Fig.~\ref{FigureC2}(b) shows that, like $\left(Z - Z_\text{iso}^N\right) N$, $GN$ also shows finite-size scaling as a function of $p^{1/2}N$ for different system sizes and pressures. Note that the slight $N$-dependence for large pressure in Fig.~\ref{FigureA}(b) completely vanishes when the data is scaled (Fig.~\ref{FigureC2}(b)). This is a result of the non-trivial scaling function at intermediate $p^{1/2}N$.

The plateaus at low $p^{1/2} N$ in the scaling functions for $\left(Z - Z_\text{iso}^N\right)N$ and $GN$ result from the fact that $Z_\text{iso}^N$ contacts per particle are not enough for the system to have a positive bulk modulus -- one additional contact is required.  These plateaus can be subtracted off in order to study the system-size dependence in greater detail. In this case, Fig.~\ref{FigureC2}(c) shows that $Z \rightarrow Z_B^N$ at low pressures, confirming Eq.~\eqref{Z_approaches_ZBN}.  Importantly,  as we asserted above, \emph{no} properly minimized configurations are observed that satisfy both $Z<Z_B^N$ and $B>0$.

Note that $\left(Z - Z_B^N\right) N$, like $\left(Z- Z_\text{iso}^N\right)N$, collapses onto a single curve when plotted versus $p^{1/2}N$ (Fig.~\ref{FigureC2}(c)).  There is, however, a crossover to $Z - Z_B^N \sim pN$ for $p^{1/2} N < {\cal O}(1)$ in both 2 and 3 dimensions. This scaling arises because quantities like $Z-Z_\text{iso}^N$ should only be singular at the jamming transition at $p=0$ in the thermodynamic limit; in finite systems they should be analytic around $p=0$.  Given the existence of scaling collapse, which has the form of Eq.~\eqref{Zscalingfunction}, the first two terms in the Taylor expansion of $Z-Z_\text{iso}^N$ in powers of $p$ must be
\eq{	Z-Z_\text{iso}^N \approx \frac{c_0}{N} + c_1pN,}
with constants $c_0$ and $c_1$. This is precisely what we observe in Fig.~\ref{FigureC2}(c), where $c_0=2$ reflects the extra contact at the transition. 

For the same reason, we find the same crossover in the shear modulus when we subtract the plateau value, $G_0 \sim \frac 1N$. Fig. \ref{FigureC2}(d) shows that $\left(G-G_0\right)N$ again collapses in both 2 and 3 dimensions when plotted against $p^{1/2}N$, with $G-G_0 \sim pN$ for $p^{1/2} N < {\cal O}(1)$. The qualitative similarity between $\left(Z - Z_B^N\right)N$ and $\left(G-G_0\right)N$ underscores the dependence of the shear modulus on the contact number. Indeed, we find that to a very good approximation, $GN$ is a pure power law in $\left(Z - Z_\text{iso}^N\right)N$ (Fig.~\ref{FigureD}), in accord with recent results of Dagois-Bohy, et al.~\cite{DagoisBohy:2012we}. 


We have also studied the finite-size scaling of the bulk modulus, $B$, which scales as $p^0$ for harmonic repulsions.  Therefore, $B$ approaches a constant value, $B_0$, as $p\rightarrow 0$.  As with the coordination number and shear modulus, we subtracted off the plateau value to study $B-B_0$.  The quantity $B-B_0$ is necessarily quite sensitive to $B_0$, which is large, in contrast to $G_0$, which is of order $1/N$.  Our results are consistent with $\left(B-B_0\right)N$ collapsing in both 2 and 3 dimensions as a function of $p^{1/2}N$ with the same asymptotic behavior as $\left(Z - Z_B^N\right)N$ and $\left(G-G_{0}\right)N$.

{\it Discussion.} We have argued that an appropriate definition of jamming is that a system can support an external stress.  One could either restrict the definition to a compressive stress, requiring $B>0$, or to any stress, requiring $B>0$ and $G>0$.
If one requires $B>0$, then sphere packings require one additional contact in the entire system, beyond the number calculated for the isostatic condition, in order to become jammed.  If one requires both $B>0$ and $G>0$, then $d(d+1)/2$ additional contacts are required.


Our results provide a simple interpretation of the results of Moukarzel~\cite{Moukarzel:2012fg}, who found that the elastic moduli vanish in the large $N$ limit for random networks with $Z=4$ in $d=2$.  Comparing $Z=4$ to $Z_\text{iso}^N$ (Eq.~\ref{Ziso_N_def}), we see that $Z>Z_\text{iso}^N$ so that $Z - Z_\text{iso}^N=4/N$.  For random spring networks, the bulk and shear moduli scale with $Z - Z_\text{iso}^N$, implying that $B$ and $G$ both scale as $1/N$ for $Z=4$.  Thus, our constraint counting arguments imply that $B$ and $G$ should vanish as $1/N$ as $N \rightarrow \infty$, consistent with Moukarzel's results.

We find that $Z - Z_\text{iso}^N$, $Z - Z_B^N$ and $G$ are analytic around the jamming transition in finite systems and exhibit finite-size scaling collapse, a defining characteristic of phase transitions.  These results cannot be understood from constraint counting alone, and provide direct evidence that the jamming transition is a phase transition.


The finite-size scaling that we observe depends on the total number of particles, $N$, rather than on the system length, $L$, in both 2 and 3 dimensions. For first-order transitions, quantities exhibit scaling collapse with $N\sim L^d$, the number of particles in the system, not with $L$, the linear size of the system~\cite{Fisher:1982uw}.  For second-order transitions in systems at or above the upper critical dimension, finite-size scaling also leads to collapse with $N$~\cite{BINDER:1985vl,Dillmann:1998ty}.  Earlier observations that the exponents do not depend on dimension in $d=2$ and $3$~\cite{Liu:2010jx} and an Imry-Ma-type argument of Wyart~\cite{Wyart:2005vu} both suggest that the jamming transition has an upper critical dimension of 2.  Our result that quantities exhibit scaling collapse as a function of $p^{1/2}N$ is therefore consistent with both the first- and mean-field second-order character of the jamming transition.  

We thank Brooks Harris, Tom Lubensky, Anton Souslov, Brian Tighe, Martin van Hecke, Peter Young and Zorana Zeravcic  for important discussions.  This research was supported by the U.S. Department of Energy, Office of Basic Energy Sciences, Division of Materials Sciences and Engineering under Awards DE-FG02-05ER46199 (AJL) and DE-FG02-03ER46088 (SRN).  CPG was supported by the NSF through a Graduate Research Fellowship.

\end{document}